\documentclass[journal,draftclsnofoot,onecolumn,12pt]{IEEEtran}
\usepackage{booktabs}

\usepackage{amssymb}
\usepackage{amsbsy}
\usepackage{amsmath}
\usepackage{bm}
\usepackage{verbatim}
\usepackage{cite}
\usepackage{mathrsfs}
\usepackage{amsfonts}
\usepackage{graphicx}
\usepackage[tight,footnotesize]{subfigure}
\usepackage[10pt]{moresize}
\usepackage{array}
\usepackage{color}
\usepackage{epsfig}
\usepackage{stfloats}
\usepackage{balance}

\usepackage{graphicx}
\usepackage{epstopdf}

\usepackage[noend]{algpseudocode}

\usepackage{algorithmicx,algorithm}
\usepackage{algpseudocode}

\usepackage{diagbox}
\usepackage{makecell}

\usepackage{cite}
\usepackage{cases}
\usepackage{setspace}

\newcommand{\subparagraph}{}
\usepackage{titlesec}

\titlespacing{\section}{0pt}{1.5ex plus .0ex minus .0ex}{1.5ex plus .0ex}
\titlespacing{\subsection}{0pt}{1ex plus .0ex minus .0ex}{1.0ex plus .0ex}
\titlespacing{\subsubsection}{0pt}{1.0ex plus .0ex minus .0ex}{1.0ex plus .0ex}
\usepackage{amsmath} 
\allowdisplaybreaks[4]

\begin{document}

\title{Performance of A Reconfigurable Intelligent Surface Controlled by An Illegitimate User}

\author{Xinwu Chen, ~Shuai Han,  and Hai Jiang
\thanks{Xinwu Chen and Shuai Han are with the School of Electronics and Information Engineering, Harbin Institute of Technology, Harbin 150001, China (e-mail: cxw808@qq.com; hanshuai@hit.edu.cn).}
\thanks{Hai Jiang is with the Department of Electrical and Computer Engineering, University of Alberta, Edmonton, AB T6G 1H9, Canada (e-mail: hai1@ualberta.ca).}

}



\maketitle

\begin{abstract}
In this paper, we investigate how an illegitimate user can use a reconfigurable intelligent surface (RIS) to maximally degrade the secrecy performance of the legitimate system. We first consider the case when the illegitimate user performs eavesdropping only. A problem is formulated to minimize the system secrecy rate by optimizing the phase shifts of the RIS elements. We also consider the case when the illegitimate user works with full-duplex technology and can perform eavesdropping and jamming simultaneously through the RIS. A problem is formulated to minimize the system secrecy rate by optimizing the jamming power and the phase shifts of the RIS elements. Although the two formulated problems are nonconvex, they are optimally solved in this paper. Our results show that good eavesdropping
performance can be achieved by placing the RIS at an appropriate
position. Further, jamming plus eavesdropping
can largely reduce the secrecy rate of the legitimate system. 
\end{abstract}


\IEEEpeerreviewmaketitle

\section{Introduction}
%
As a near-passive technology, a reconfigurable intelligent surface (RIS) \cite{di2020smart,wu2019towards} has many reflective elements that can be with low cost. By controlling these elements, the electromagnetic wave paths from the outside environment into the RIS can be phase-shifted. On the other hand, since RIS is a passive device, the amplitude of the electromagnetic waves cannot be adjusted beyond the original amplitude. Rather, the amplitude can be reduced to eliminate the unwanted electromagnetic energy \cite{wu2019beamforming}.

Recently, a lot of attention has been paid to RIS-assisted physical layer security \cite{yang2020secrecy,guan2020intelligent,chu2020secrecy,yang2020deep}. In \cite{yang2020secrecy}, it was shown that RIS is effective to enhance physical layer security. Guan \emph{et al.} \cite{guan2020intelligent} showed that the security of RIS communications was improved with the help of artificial noise. Chu \emph{et al.} \cite{chu2020secrecy} used an RIS-assisted multiple-input multiple-output secrecy system to improve the secrecy performance and used artificial noise to reduce the eavesdropper's reception performance. In \cite{yang2020deep}, base station beamforming and reflected beamforming are jointly designed to improve the system secrecy rate.

However, in the literature, there is no research effort for the possible impact on wireless communication security when an illegitimate receiver is aided by RIS. Therefore, in this paper, we propose to investigate how an RIS-assisted illegitimate receiver can maximally degrade the secrecy performance of a legitimate receiver. We further investigate the case when the illegitimate receiver has full duplex technology and can send jamming signals to further deteriorate the legitimate user's performance. Our research can help the legitimate system to evaluate its worst secrecy situation. The main contributions of this paper are as follows.

\begin{itemize}	
	\item[$\bullet$] We propose to evaluate wireless secrecy performance from a new perspective: what if an RIS is controlled by an illegitimate receiver? Specifically, we consider that an illegitimate receiver uses the help of an RIS to perform its eavesdropping. The system is then extended to the case when the illegitimate receiver uses full duplex technology to do both eavesdropping and jamming.
	\item[$\bullet$] For the eavesdropping-only system and eavesdropping-plus-jamming system, we formulate optimization problems to minimize the secrecy rate of the legitimate receiver. The formulated problems are non-convex. The non-convex problems are solved by using some math manipulations.
	\item[$\bullet$] Our simulation results show that good eavesdropping performance can be achieved by placing the RIS at an appropriate position. Further, jamming plus eavesdropping can largely reduce the secrecy rate of the legitimate system.	
\end{itemize}
The rest of this paper is organized as follows: Section \uppercase\expandafter{\romannumeral2} gives the system model with a RIS-aided illegitimate user, and formulate two problems when the illegitimate user performs eavesdropping only or performs eavesdropping plus jamming. Section \uppercase\expandafter{\romannumeral3} gives our algorithms to solve the formulated problems. Section \uppercase\expandafter{\romannumeral4} presents performance evaluation of our algorithms. Finally, we give the conclusion in Section \uppercase\expandafter{\romannumeral5}.
\section{SYSTEM MODEL AND PROBLEM FORMULATION}
\subsection{System Model}
\begin{figure}[!h]
	\centering\includegraphics[width=4in]{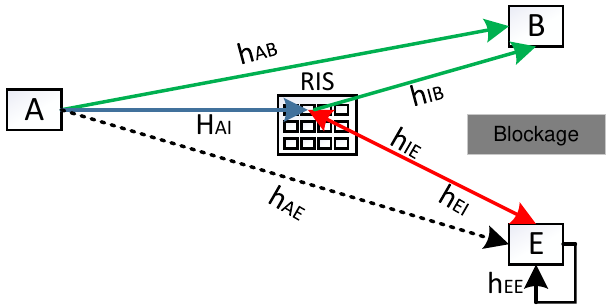}
	\caption{System model}
	\label{model}
\end{figure}
Fig.~\ref{model} shows our system model.  We consider a communication scenario in which a multi-antenna message sender (A) sends a confidential message to a single-antenna legitimate receiver (B) with transmit power ${{P}_{T}}$. A has $M$ antennas. In addition, there exists an illegitimate receiver (E) with a single antenna, which tries to perform illegitimate information eavesdropping and/or jamming by using the help of an RIS that has $N$ reflecting elements. There exists a blockage between E and B. The baseband equivalent channels from A to the RIS, to B and to E are denoted as ${{\mathbf{H}}_{AI}}\in {{\mathbb{C}}^{N\times M}}$, ${{\mathbf{h}}_{AB}}\in {{\mathbb{C}}^{1\times M}}$, ${{\mathbf{h}}_{AE}}\in {{\mathbb{C}}^{1\times M}}$, respectively. The baseband equivalent channels from the RIS to B and to E are denoted by ${{\mathbf{h}}_{IB}}\in {{\mathbb{C}}^{1\times N}}$ and ${{\mathbf{h}}_{IE}}\in {{\mathbb{C}}^{1\times N}}$, respectively. Further, if E has full duplex technology, it can eavesdrop and simultaneously interfere with E by transmitting jamming signals to B with the RIS's assistance. The communication channel from E to the RIS is denoted as ${{\mathbf{h}}_{EI}}\in {{\mathbb{C}}^{N\times 1}}$. The self-interfering channel of E is denoted as ${{\mathbf{h}}_{EE}}\in {{\mathbb{C}}^{1\times 1}}$. Let $\mathbf{\Theta }=diag({{\alpha }_{1}}\exp (j{{\theta }_{1}}),{{\alpha }_{2}}\exp (j{{\theta }_{2}}),\ldots ,{{\alpha }_{N}}\exp (j{{\theta }_{N}}))$ be the reflecting coefficients of the RIS, where ${{\alpha }_{n}}\in [0,1]$ is the amplitude of  the ${n}$th RIS element and ${{\theta }_{n}}\in [0,2\pi )$ is the phase shift of the ${n}$th RIS element. Here, in order to obtain the maximum reflecting power gain, we set the amplitude of RIS as ${{\alpha }_{n}}=1,n\in N$. 

Two cases are investigated, as follows.

Case 1: E performs passive eavesdropping only. When A transmits, the received signal at B and E are given by
\begin{equation}\label{yb1}
	\begin{aligned}[b]
		{{y}_{B\text{1}}}=({{\mathbf{h}}_{AB}}+\mathbf{h}_{IB}\mathbf{\Theta }{{\mathbf{H}}_{AI}}){{\mathbf{w}}_{B}}x+{{n}_{B}}
	\end{aligned}
\end{equation}
and
\begin{equation}\label{ye1}
	\begin{aligned}[b]
		{{y}_{E\text{1}}}=({{\mathbf{h}}_{AE}}+\mathbf{h}_{IE}\mathbf{\Theta }{{\mathbf{H}}_{AI}}){{\mathbf{w}}_{B}}x+{{n}_{E}},
	\end{aligned}
\end{equation}
respectively, where ${{\mathbf{w}}_{B}}\in {{\mathbb{C}}^{M\times 1}}$ denotes A's beamforming vector, $x\sim CN(0,1)$ denotes A's information, and ${{n}_{B}}\sim CN(0,\sigma _{B}^{2})$ and ${{n}_{E}}\sim CN(0,\sigma _{E}^{2})$ are complex additive white Gaussian noise at B and E, respectively.
Then the signal-to-interference-plus-noise ratio (SINR) at B and E are given as
\begin{equation}\label{rb1}
	\begin{aligned}[b]
		{{\gamma }_{B1}}=\frac{{{\left| ({{\mathbf{h}}_{AB}}+\mathbf{h}_{IB}\mathbf{\Theta }{{\mathbf{H}}_{AI}}){{\mathbf{w}}_{B}} \right|}^{\text{2}}}}{\sigma _{B}^{2}}
	\end{aligned}
\end{equation}
and
\begin{equation}\label{re1}
	\begin{aligned}[b]
		{{\gamma }_{E1}}=\frac{{{\left| ({{\mathbf{h}}_{AE}}+\mathbf{h}_{IE}\mathbf{\Theta }{{\mathbf{H}}_{AI}}){{\mathbf{w}}_{B}} \right|}^{\text{2}}}}{\sigma _{E}^{2}},
	\end{aligned}
\end{equation}
respectively.
Thus, the secrecy rate from A to B can be expressed as
\begin{equation}\label{R1}
	\begin{aligned}[b]
		{{R}_{\sec 1}}\text{= }\!\![\!\!\text{ }{{\log }_{2}}(1+{{\gamma }_{B1}})-{{\log }_{2}}(1+{{\gamma }_{E1}}){{]}^{+}},
	\end{aligned}
\end{equation}
where ${{\text{ }\!\![\!\!\text{ }k]}^{+}}\triangleq \max (0,k)$.

Case 2: E adopts a full-duplex mode, and thus, operates and controls the RIS for passive eavesdropping and active jamming. The signal reception at B and E are shown as
\begin{equation}\label{yb2}
	\begin{aligned}[b]
		{{y}_{B2}}=({{\mathbf{h}}_{AB}}+\mathbf{h}_{IB}\mathbf{\Theta }{{\mathbf{H}}_{AI}}){{\mathbf{w}}_{B}}x+(\mathbf{h}_{IB}\mathbf{\Theta }{{\mathbf{h}}_{EI}})z+{{n}_{B}}
	\end{aligned}
\end{equation}
and
\begin{equation}\label{ye2}
	\begin{aligned}[b]
	{{y}_{E\text{2}}}&=({{\mathbf{h}}_{AE}}+\mathbf{h}_{IE}\mathbf{\Theta }{{\mathbf{H}}_{AI}}){{\mathbf{w}}_{B}}x \\ &+(\sqrt{\rho }{{\mathbf{h}}_{EE}}+\mathbf{h}_{IE}\mathbf{\Theta }{{\mathbf{h}}_{EI}})z+{{n}_{E}},
	\end{aligned}
\end{equation}
respectively, where $z\sim CN(0,1)$ denote the jamming signal, and $\rho$ denotes the suppression ratio for the self-interference channel with full duplex at E.

Therefore, the SINR at B and E are given as,
\begin{equation}\label{rb2}
	\begin{aligned}[b]
		{{\gamma }_{B2}}=\frac{{{\left| ({{\mathbf{h}}_{AB}}+\mathbf{h}_{IB}\mathbf{\Theta }{{\mathbf{H}}_{AI}}){{\mathbf{w}}_{B}} \right|}^{\text{2}}}}{{{P}_{j}}{{\left| \mathbf{h}_{IB}\mathbf{\Theta }{{\mathbf{h}}_{EI}} \right|}^{2}}+\sigma _{B}^{2}}
	\end{aligned}
\end{equation}
and
\begin{equation}\label{re2}
	\begin{aligned}[b]
		{{\gamma }_{E2}}=\frac{{{\left| ({{\mathbf{h}}_{AE}}+\mathbf{h}_{IE}\mathbf{\Theta }{{\mathbf{H}}_{AI}}){{\mathbf{w}}_{B}} \right|}^{\text{2}}}}{{{P}_{j}}{{\left| \sqrt{\rho }{{\mathbf{h}}_{EE}}+\mathbf{h}_{IE}\mathbf{\Theta }{{\mathbf{h}}_{EI}} \right|}^{2}}+\sigma _{E}^{2}},
	\end{aligned}
\end{equation}
respectively, where ${{P}_{j}}$ represents the power of E when transmitting jamming signals. The secrecy rate from A to B in this case can be given as
\begin{equation}\label{R3}
	\begin{aligned}[b]
		{{R}_{\sec 2}}\text{= }\!\![\!\!\text{ }{{\log }_{2}}(1+{{\gamma }_{B2}})-{{\log }_{2}}(1+{{\gamma }_{E2}}){{]}^{+}}.
	\end{aligned}
\end{equation}
\subsection{Problem Formulation}
We aim to minimize the achievable secrecy rate of B via optimizing the reflecting coefficients of RIS in Case 1, or jointly optimizing the reflecting coefficients of the RIS and  the transmit power of E in Case 2. Accordingly, the optimization problems for the two cases are formulated, respectively, as
\begin{subequations}\label{P1}
	\begin{align}
		(\text{P1}): \underset{\mathbf{\Theta}}{\mathop{\min }}\, &
		{{R}_{\sec 1}} \\
		\mbox{s.t.}\
		{} & \left| \exp (j{{\theta }_{n}}) \right|\text{=1,}n\in N		
	\end{align}
\end{subequations}
and
\begin{subequations}\label{P2}
	\begin{align}
			(\text{P2}):  \underset{\mathbf{\Theta},{{P}_{j}}}{\mathop{\min }}\, & {{R}_{\sec 2}} \\
		 \mbox{s.t.}\ & 0\le {{P}_{j}}\le {{P}_{D}}  \\
		 {} & \left| \exp (j{{\theta }_{n}}) \right|\text{=1,}~n\in N,
	\end{align}
\end{subequations}
where ${{P}_{D}}$ represents maximum transmit power of E.

It is generally difficult to optimally solve Problem P1 and P2, as they are non-convex problem and they are subject to the unit-norm constraint.
In the next section, we will propose effective algorithms to optimally solve the two problems.

\section{Proposed Algorithms for Problem P1 and P2}
\subsection{Solution to Problem P1}\label{sec:P1_sol}
From (\ref{R1}), Problem P1 is equivalent to the following problem
\begin{subequations}\label{P1.1}
	\begin{align}
		(\text{P1.1}): \underset{\mathbf{\Theta}}{\mathop{\min }}\, &
		\frac{\frac{\text{1}}{\sigma _{B}^{2}}{{\left| ({{\mathbf{h}}_{AB}}+\mathbf{h}_{IB}\mathbf{\Theta }{{\mathbf{H}}_{AI}}){{\mathbf{w}}_{B}} \right|}^{\text{2}}}\text{+1}}{\frac{\text{1}}{\sigma _{E}^{2}}{{\left| ({{\mathbf{h}}_{AE}}+\mathbf{h}_{IE}\mathbf{\Theta }{{\mathbf{H}}_{AI}}){{\mathbf{w}}_{B}} \right|}^{\text{2}}}\text{+1}} \\
		\mbox{s.t.}\
		{} & \left| \exp (j{{\theta }_{n}}) \right|\text{=1,}~n\in N.		
	\end{align}
\end{subequations}

Let ${{\mathbf{v}}^{H}}=[{{v}_{1}},{{v}_{2}},{{v}_{3}},\ldots ,{{v}_{N}},1]$, where ${{v}_{n}}=\alpha \exp (j{{\theta }_{n}}),\forall n$. Then, we have
\begin{equation}\label{b1}
	\begin{aligned}[b]
		({{\mathbf{h}}_{AB}}+\mathbf{h}_{IB}\mathbf{\Theta }{{\mathbf{H}}_{AI}})=\mathbf{v}^{H}{{\mathbf{f}}_{B}},
	\end{aligned}
\end{equation}
\begin{equation}\label{e1}
	\begin{aligned}[b]
		({{\mathbf{h}}_{AE}}+\mathbf{h}_{IE}\mathbf{\Theta }{{\mathbf{H}}_{AI}})=\mathbf{v}^{H}{{\mathbf{f}}_{E}},
	\end{aligned}
\end{equation}
where
\begin{equation}\label{fb}
	\begin{aligned}[b]
		{{\mathbf{f}}_{B}}=\left[ \begin{matrix}
			diag(\mathbf{h}_{IB}){{\mathbf{H}}_{AI}}  \\
			{{\mathbf{h}}_{AB}}  \\
		\end{matrix} \right],
	\end{aligned}
\end{equation}
\begin{equation}\label{fe}
	\begin{aligned}[b]
		{{\mathbf{f}}_{E}}=\left[ \begin{matrix}
			diag(\mathbf{h}_{IE}){{\mathbf{H}}_{AI}}  \\
			{{\mathbf{h}}_{AE}}  \\
		\end{matrix} \right].
	\end{aligned}
\end{equation}

Define matrix ${{\mathbf{W}}_{B}}\triangleq {{\mathbf{w}}_{B}}\mathbf{w}_{B}^{H}$ and $\mathbf{V}\triangleq \mathbf{v}{{\mathbf{v}}^{H}}$, which satisfy ${{\mathbf{W}}_{B}}\succeq \mathbf{0}$, ${{\mathbf{V}}}\succeq \mathbf{0}$, $\text{rank}({{\mathbf{W}}_{B}})=1$, and $\text{rank}({{\mathbf{V}}})=1$. Using (\ref{b1}) and (\ref{e1}), Problem P1.1 is equivalent to the following problem:
\begin{subequations}\label{P1.2}
	\begin{align}
		(\text{P1.2}): \underset{\mathbf{V}\succeq0}{\mathop{\min }}\, &
		\frac{\frac{\text{1}}{\sigma _{B}^{2}}\text{Tr}\text{(}{{\mathbf{f}}_{B}}{{\mathbf{W}}_{B}}\mathbf{f}_{B}^{H}\mathbf{V}\text{)+1}}{\frac{\text{1}}{\sigma _{E}^{2}}\text{Tr}\text{(}{{\mathbf{f}}_{E}}{{\mathbf{W}}_{B}}\mathbf{f}_{E}^{H}\mathbf{V}\text{)+1}} \\
		\mbox{s.t.}\
		{} & \text{Tr}({{\mathbf{E}}_{n}}\mathbf{V})=1,~n\in\{1,2,3,...,N+1\}		
	\end{align}
\end{subequations}
where ${\mathbf{E}}_{n}$ is an $(N+1)\times(N+1)$ matrix in which only one element (i.e., the element at the $n$th row and $n$th column) is 1 and all other elements are 0.
%
Problem P1.2 is still a non-convex problem. We apply the Charnes-Cooper transform to change the fractional structure of the problem \cite{cui2019secure}, i.e., introduce an auxiliary variable to decouple the numerator and denominator as follows. A detailed description of the Charnes-Cooper transform is given in \cite{shen2018fractional}.

We set ${{\mu }_{1}}=1/(\frac{\text{1}}{\sigma _{E}^{2}}\text{Tr}\text{(}{{\mathbf{f}}_{E}}{{\mathbf{W}}_{B}}\mathbf{f}_{E}^{H}\mathbf{V}\text{)+1})$ and $\mathbf{X}={{\mu }_{1}}\mathbf{V}$. Then Problem P1.2 can be transformed into a non-fractional structural form as
\begin{subequations}\label{P1.3}
	\begin{align}
		(\text{P1.3}): \underset{{{\mu}_{2}\succeq 0},\mathbf{X}\succeq0}{\mathop{\min }}\, &
		\frac{\text{1}}{\sigma _{B}^{2}}\text {Tr}\text{(}{{\mathbf{f}}_{B}}{{\mathbf{W}}_{B}}\mathbf{f}_{B}^{H}\mathbf{X}\text{)+}{{\mu }_{1}} \\
		\mbox{s.t.}\
		{} & \frac{\text{1}}{\sigma _{E}^{2}}\text {Tr}\text{(}{{\mathbf{f}}_{E}}{{\mathbf{W}}_{B}}\mathbf{f}_{E}^{H}\mathbf{X}\text{)+}{{\mu }_{1}}=1\\
		{} & \text {Tr}({{\mathbf{E}}_{n}}\mathbf{X})=\mu_{1},\forall n.		
	\end{align}
\end{subequations}

Problem P1.3 is a convex semidefinite programming (SDP) problem. Thus, its optimal solution (i.e., optimal $\mathbf{X}$) can be found by using the interior-point method or convex optimization solver (such as CVX). With optimal $\mathbf{X}$ value denoted as $\mathbf{X}^*$, we can get optimal $\mathbf{V}$ as $\mathbf{V}^*=\mathbf{X}^*/\mu_1$. Recall $\mathbf{V}\triangleq \mathbf{v}{{\mathbf{v}}^{H}}$. Accordingly, based on $\mathbf{V}^*$, we apply the standard Gaussian randomization method \cite{wu2019intelligent} to get optimal $\mathbf{v}$, which is our optimal solution to Problem P1.

\subsection{Solution to Problem P2}
Problem P2 has two different variables. So we can use the alternating optimization method to optimize it: fix one variable and then optimize the other variable, alternate between the two variables until the solution converges.

Similar to Section \ref{sec:P1_sol}, we can have
\begin{equation}\label{b2}
	\begin{aligned}[b]
		\mathbf{h}_{IB}\mathbf{\Theta }{{\mathbf{h}}_{EI}}=\mathbf{v}^{H}{{\mathbf{f}}_{B1}},
	\end{aligned}
\end{equation}
\begin{equation}\label{e2}
	\begin{aligned}[b]
		\sqrt{\rho }{{\mathbf{h}}_{EE}}+\mathbf{h}_{IE}\mathbf{\Theta }{{\mathbf{h}}_{EI}}=\mathbf{v}^{H}{{\mathbf{f}}_{E1}},
	\end{aligned}
\end{equation}
where
\begin{equation}\label{fb1}
	\begin{aligned}[b]
		{{\mathbf{f}}_{B1}}=\left[ \begin{matrix}
			diag(\mathbf{h}_{IB}){{\mathbf{h}}_{EI}}  \\
			\mathbf{0}  \\
		\end{matrix} \right],
	\end{aligned}
\end{equation}
\begin{equation}\label{fe2}
	\begin{aligned}[b]
		{{\mathbf{f}}_{E1}}=\left[ \begin{matrix}
			diag(\mathbf{h}_{IE}){{\mathbf{h}}_{EI}}  \\
			\sqrt{\rho }{{\mathbf{h}}_{EE}}  \\
		\end{matrix} \right].
	\end{aligned}
\end{equation}
Then we can transform Problem P2 to
\begin{subequations}\label{P2.1}
	\begin{align}
		(\text{P2.1}): \underset{\mathbf{V},{{P}_{j}}}{\mathop{\min }}\, & {R_{b}}-{R_{e}} \\
		\mbox{s.t.}\  & 0\le {{P}_{j}}\le {{P}_{D}},  \\
		{} & {{\mathbf{V}}_{n,n}}=1,n=1,\ldots ,N,N+1,  \\
		{} & \mathbf{V}\succeq 0
	\end{align}
\end{subequations}
where
\begin{equation}\label{Rb_ini}
	\begin{aligned}[b]
		R_{b}={{\log }_{2}}(1+\frac{\text{Tr(}{{\mathbf{f}}_{B}}{{\mathbf{W}}_{B}}\mathbf{f}_{B}^{H}\mathbf{V}\text{)}}{{{P}_{j}}\text{Tr}({{\mathbf{f}}_{B1}}\mathbf{f}_{B1}^{H}\mathbf{V})+\sigma _{B}^{2}}),
	\end{aligned}
\end{equation}
\begin{equation}\label{Re}
	\begin{aligned}[b]
		R_{e}={{\log }_{2}}(1+\frac{\text{Tr(}{{\mathbf{f}}_{E}}{{\mathbf{W}}_{B}}\mathbf{f}_{E}^{H}\mathbf{V}\text{)}}{{{P}_{j}}\text{Tr}({{\mathbf{f}}_{E1}}\mathbf{f}_{E1}^{H}\mathbf{V})+\sigma _{E}^{2}}).
	\end{aligned}
\end{equation}

Problem P2.1 is still a non-trivial problem. To solve it, we introduce the following statement, which is Lemma 1 in \cite{li2013transmit}.

{\bf Statement 1}: Consider a specific value $x>0$ and define a function of $t$ as $\varphi (t)=tx-\ln t-1$. For minimization problem $\underset{t>0}{\mathop{\min}}\,\varphi (t)$, its minimal objective function is $\ln x$, which happens when $t=1/x.$
%
%

According to the above statement, ${R}_{b}$ in (\ref{Rb_ini}) and $R_{e}$ in (\ref{Re}) can be expressed as
\begin{equation}\label{Rb}
	\begin{aligned}[b]
		{{R}_{b}} & =\frac{1}{\ln 2}\underset{{{\lambda }_{B}}>0}{\mathop{\min }}\,{{\varphi }_{B}}(\mathbf{V},{{\lambda }_{B}},P_{j}),
	\end{aligned}
\end{equation}
\begin{equation}\label{Re1}
	\begin{aligned}[b]
		-{{R}_{e}}\text{ }=\frac{1}{\ln 2}\underset{{{\lambda }_{E}}>0}{\mathop{\min }}\,{{\varphi }_{E}}(\mathbf{V},{{\lambda }_{E}},{{P}_{j}}),
	\end{aligned}
\end{equation}
where
\begin{equation}\label{faiB}
	\begin{aligned}[b]
		&{{\varphi }_{B}}(\mathbf{V},{{\lambda }_{B}},P_{j}) \\& ={{\lambda }_{B}}(\text{Tr(}{{\mathbf{f}}_{B}}\mathbf{W}_{B}\mathbf{f}_{B}^{H}\mathbf{V}\text{)+}{{P}_{j}}\text{Tr}({{\mathbf{f}}_{B1}}\mathbf{f}_{B1}^{H}\mathbf{V})+\sigma _{B}^{2}) \\
		&-\ln({{P}_{j}}\text{Tr}({{\mathbf{f}}_{B1}}\mathbf{f}_{B1}^{H}\mathbf{V})+\sigma _{B}^{2})-\ln ({{\lambda }_{B}})-1,
	\end{aligned}
\end{equation}
\begin{equation}\label{faiE}
	\begin{aligned}[b]
		&{{\varphi }_{E}}(\mathbf{V},{{\lambda }_{E}},{{P}_{j}})\\&=-\ln (\text{Tr(}{{\mathbf{f}}_{E}}{{\mathbf{W}}_{B}}\mathbf{f}_{E}^{H}\mathbf{V}\text{)}+{{P}_{j}}\text{Tr}({{\mathbf{f}}_{E1}}\mathbf{f}_{E1}^{H}\mathbf{V})+\sigma _{E}^{2})\\
		&+{{\lambda }_{E}}({{P}_{j}}\text{Tr}({{\mathbf{f}}_{E1}}\mathbf{f}_{E1}^{H}\mathbf{V})+\sigma _{E}^{2})-\ln ({{\lambda }_{E}})-1.
	\end{aligned}
\end{equation}
Next, we solve Problem P2.1 by using alternating optimization as follows.

We fix $\mathbf{V}$ as an initial value $\mathbf{V}^\dag$ and fix $P_{j}$ as an initial value $P_{j}^\dag$ , and obtain the optimal solution for (\ref{faiB}) and (\ref{faiE}) as
\begin{equation}\label{lambdaB}
	\begin{aligned}[b]
		\lambda _{B}^{*}={{(\text{Tr(}{{\mathbf{f}}_{B}}{{\mathbf{W}}_{B}}\mathbf{f}_{B}^{H}\mathbf{V}^\dag\text{)+}{{P}_{j}}\text{Tr}({{\mathbf{f}}_{B1}}\mathbf{f}_{B1}^{H}\mathbf{V}^\dag)+\sigma _{B}^{2})}^{-1}},
	\end{aligned}
\end{equation}
\begin{equation}\label{lambdaE}
	\begin{aligned}[b]
		\lambda _{E}^{*}={{({{P}_{j}^\dag}\text{Tr}({{\mathbf{f}}_{E1}}\mathbf{f}_{E1}^{H}\mathbf{V}^\dag)+\sigma _{E}^{2})}^{-1}}.
	\end{aligned}
\end{equation}

We still fix  $P_{j}$ as $P_{j}^\dag$, and fix $\lambda _{B}$ as $\lambda _{B}^*$ and fix $\lambda _{E}$ as $\lambda _{E}^*$. Then Problem P2.1 can be transformed to
\begin{subequations}\label{P2.1.1}
	\begin{align}
		(\text{P2.1.1}): \underset{\mathbf{V}}{\mathop{\min}}\, & ~~~{{\varphi }_{B}}(\mathbf{V},{{\lambda }_{B}^{*}}, P_{j}^\dag)+ {{\varphi }_{E}}(\mathbf{V},{{\lambda }_{E}^{*}}, P_{j}^\dag)   \\
		\mbox{s.t.}\
		{} & {{\mathbf{V}}_{n,n}}=1,n=1,\ldots ,N,N+1,  \\
		{} & \mathbf{V}\succeq 0.
	\end{align}
\end{subequations}

Problem P2.1.1 is a convex problem, and its solution denoted ${\mathbf{V}}^*$ can be obtained similar to solving Problem P1.3.

Next, we fix $\mathbf{V}$ as ${\mathbf{V}}^*$, and fix $\lambda _{B}$ as $\lambda _{B}^*$ and fix $\lambda _{E}$ as $\lambda _{E}^*$. Then Problem P2.1 can be transformed to
\begin{subequations}\label{P2.1.2}
	\begin{align}
		(\text{P2.1.2}): \underset{P_{j}}{\mathop{\min }}\, & ~~~{{\varphi }_{B}}({\mathbf{V}}^*,{{\lambda }_{B}^{*}}, P_{j})+ {{\varphi }_{E}}({\mathbf{V}}^*,{{\lambda }_{E}^{*}}, P_{j})   \\
		\mbox{s.t.}\
		{} & 0\le {{P}_{j}}\le {{P}_{D}}.
	\end{align}
\end{subequations}

Problem P2.1.2 is a single-variable one-dimensional extremum problem. It can be solved directly by methods such as one-dimensional search. Denote its optimal solution as $P_j^*$.

By using the obtained ${\mathbf{V}}^*$ and $P_j^*$, we update $\lambda _{B}^{*}$ and $\lambda _{E}^{*}$ as follows.
\begin{equation}\label{lambdaB_up}
	\begin{aligned}[b]
		\lambda _{B}^{*}={{(\text{Tr(}{{\mathbf{f}}_{B}}{{\mathbf{W}}_{B}}\mathbf{f}_{B}^{H}\mathbf{V}^*\text{)+}{{P}_{j}^*}\text{Tr}({{\mathbf{f}}_{B1}}\mathbf{f}_{B1}^{H}\mathbf{V}^*)+\sigma _{B}^{2})}^{-1}}.
	\end{aligned}
\end{equation}
\begin{equation}\label{lambdaE_up}
	\begin{aligned}[b]
		\lambda _{E}^{*}={{({{P}_{j}^*}\text{Tr}({{\mathbf{f}}_{E1}}\mathbf{f}_{E1}^{H}\mathbf{V}^*)+\sigma _{E}^{2})}^{-1}}.
	\end{aligned}
\end{equation}

Then we proceed to Problem P2.1.1 again by setting $P_j^\dag = P_j^*$. The procedure is repeated until convergence.

The overall algorithm for solving Problem P2 is presented in Algorithm 1.
\begin{algorithm}[htb]
	\renewcommand{\algorithmicrequire}{\textbf{Input:}}
	\renewcommand{\algorithmicensure}{\textbf{Output:}}
	\caption{Proposed Algorithm for Problem P2}
	\label{alg1}	
	\begin{algorithmic}[1]
		\Require  $\mathbf{W}_{B}$,${{\mathbf{f}}_{B}}$,${{\mathbf{f}}_{B1}}$,$P_{D}$,${{\mathbf{f}}_{E}}$,${{\mathbf{f}}_{E1}}$.
		\Ensure $\mathbf{V}$,$P_{j}$
		\State {Initialize ${{\mathbf{V}}^{(0)}}$ and $P_j^{(0)}$}.
		\State  Set $r = 1$
		\Repeat
		\State {With given ${{\mathbf{V}}^{(r-1)}}$ and $P_{j}^{(r-1)}$,find the optimal $\lambda _{B}^{(r)}$ and $\lambda _{E}^{(r)}$ according to (\ref{lambdaB})and (\ref{lambdaE}), respectively.}
		\State {With given $\lambda _{B}^{(r)}$, $\lambda _{E}^{(r)}$ and $P_{j}^{(r-1)}$, find the optimal ${{\mathbf{V}}^{(r)}}$ by solving (\ref{P2.1.1}).}
		\State {With given $\lambda _{B}^{(r)}$, $\lambda _{E}^{(r)}$ and ${{\mathbf{V}}^{(r)}}$, find the optimal $P_{j}^{(r)}$ by sloving (\ref{P2.1.2}).}
		\State Updata $r=r+1.$
		\Until {the objective value of Problem P2 reaches convergence.}		
	\end{algorithmic}
\end{algorithm}
\section{SIMULATION RESULTS}
\begin{figure}[!h]
	\centering\includegraphics[width=4in]{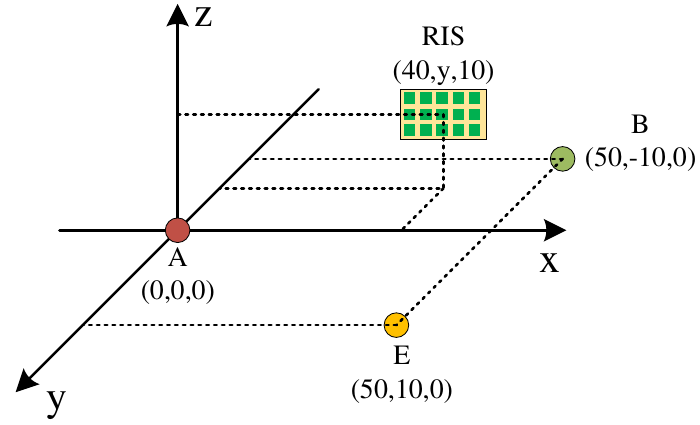}
	\caption{Simulation setup.}
	\label{simulation}
\end{figure}
Next, simulation results are provided to verify the
effectiveness of our proposed algorithms. The simulation setup is shown in Fig.~\ref{simulation}. A, B and E are located at (0, 0, 0), (50, -10, 0), and (50, 10, 0) in meter (m), respectively. A has 8 antennas, while B and E have a single antenna. The central point of the RIS is at ($40, y, 10$), where y denotes the position value of the RIS on the $y$-axis. By changing $y$, we can adjust the distance of the RIS to other devices.

For two devices $x$ and $z$, the channel between them is given as ${{h}_{x,z}}=\sqrt{{{L}_{0}}d _{x,z}^{-{{\beta }_{x,z}}}}{{g}_{x,z}}$, where ${L}_{0}=-30$dB is the large-scale fading at 1 m (reference distance), and $d _{x,z}$, ${\beta }_{x,z}$, and  ${g}_{x,z}$ are distance, path loss exponent, and small-scale fading component, respectively, between $x$ and $z$, with $x\in \{A,I\}$, $y\in \{I,E,B\}$. The path loss exponent is 4 for channels from A to B and from A to E , is 3.5 for channel from A to the RIS, and is 2.5 for channels from the RIS to B and from the RIS to E. The background noise power of the channels is -120dBW. A transmits with a transmit power level of 5 W. The maximal transmit power of E is 5 W. When E is in full-duplex mode, there is a self-interference factor of 120 dB.

\begin{figure}[!h]
	\centering\includegraphics[width=4in]{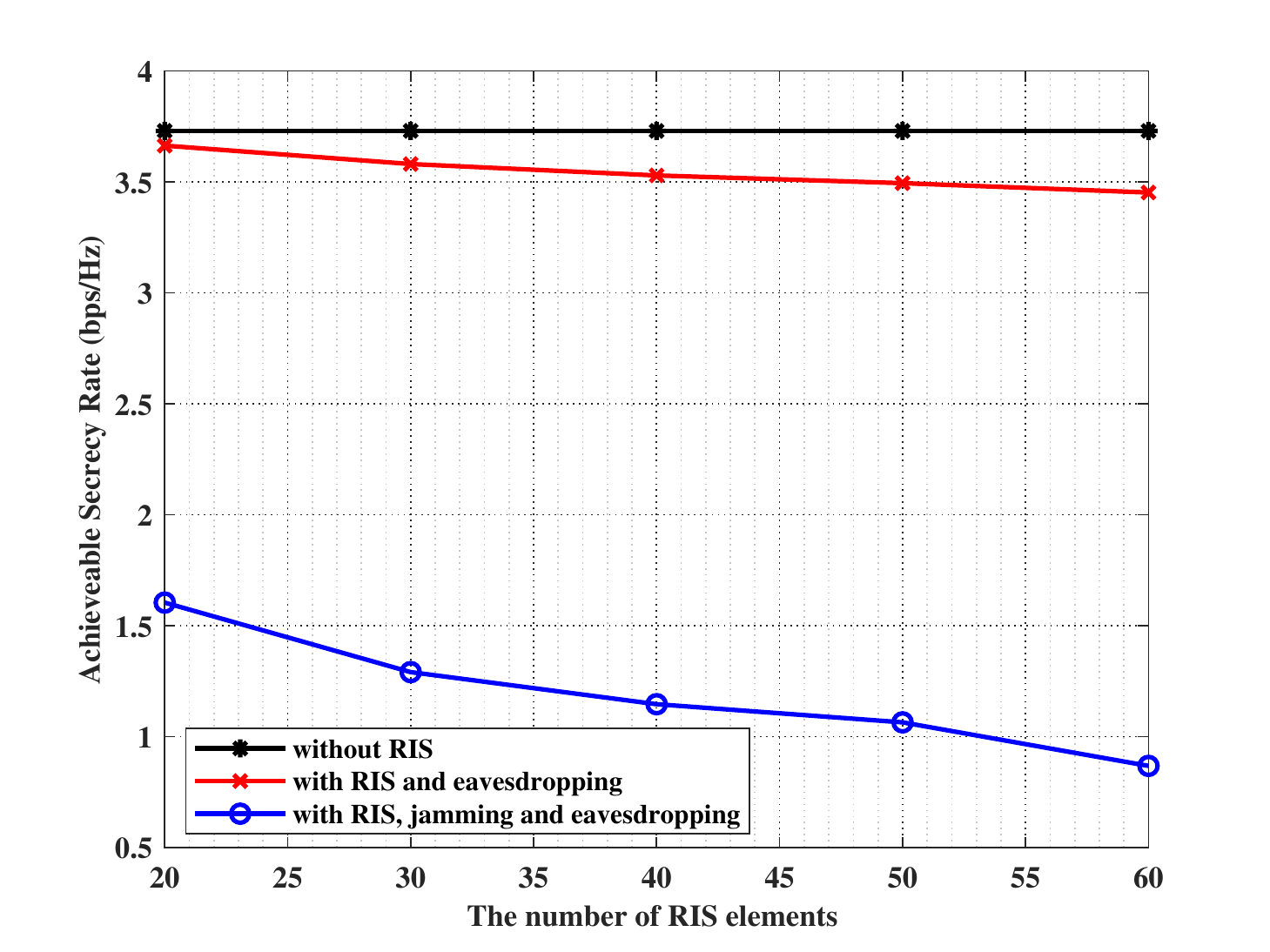}
	\caption{Achievable secrecy rate of B versus the number of RIS elements $N$.}
	\label{S-fig1}
\end{figure}

\begin{figure}[!h]
	\centering\includegraphics[width=4in]{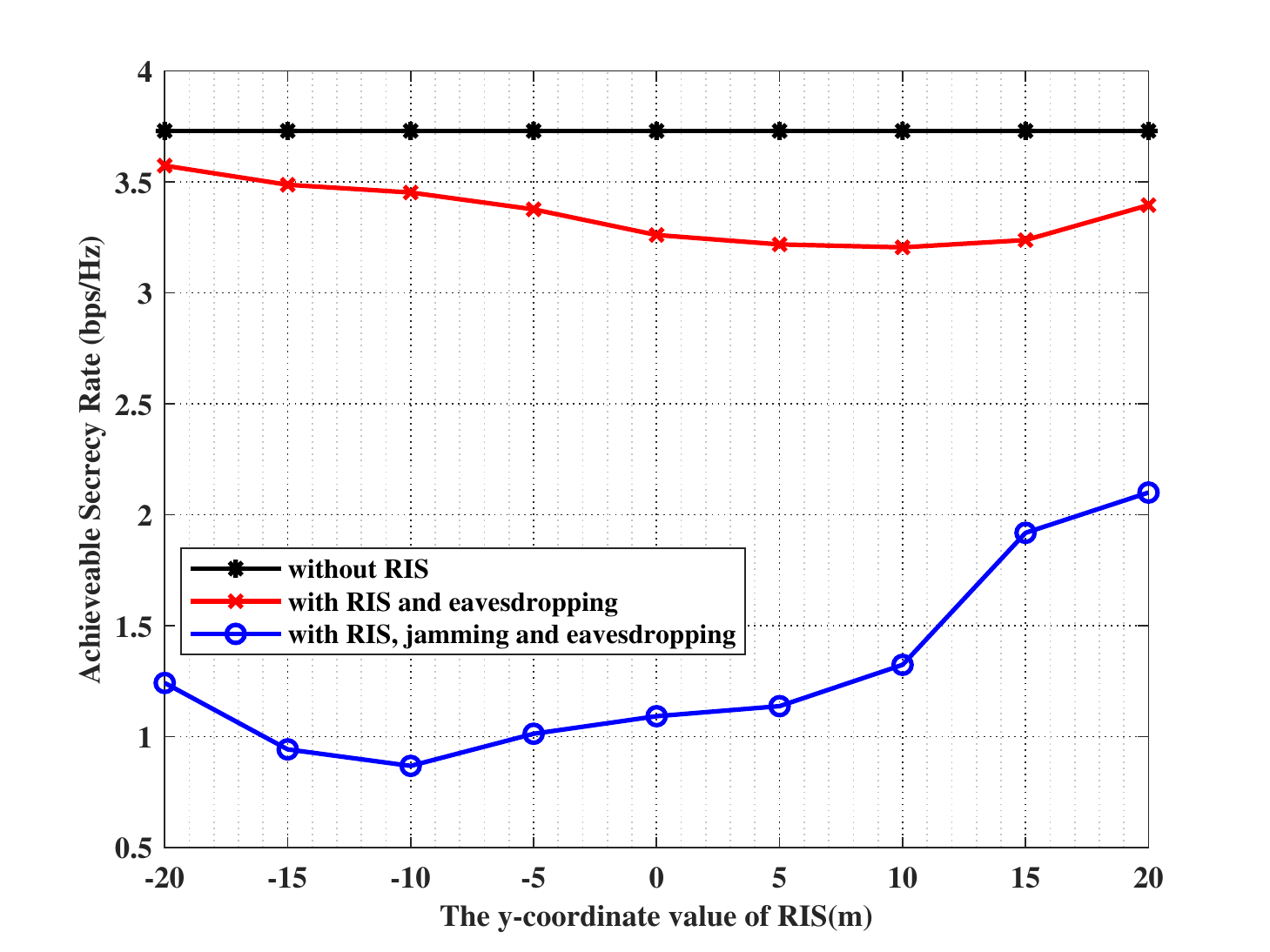}
	\caption{Achievable secrecy rate of B versus the location of RIS}
	\label{S-fig2}
\end{figure}
%
%

When the number of RIS elements $N$ changes from 20 to 60 and $y$ is fixed at -10 m, Fig.~\ref{S-fig1} shows the achievable secrecy rate of B for the case without RIS, with RIS performing eavesdropping only, and with RIS performing eavesdropping and jamming. It can be seen that for the two cases with RIS, the secrecy rate decreases as the number of RIS elements increases. It can also be seen that by using jamming, E can be much more effective in degrading the secrecy performance of the legitimate system. 

When the number of RIS elements $N$ is fixed at 60 and $y$ changes from -20 m to 20 m, Fig.~\ref{S-fig2} depicts the achievable secrecy rate of B for the three cases. When RIS is used for eavesdropping only, the best eavesdropping performance is achieved when the RIS is close to E. When RIS is used to perform both eavesdropping and jamming, the secrecy rate of the legitimate system is largely reduced. It is seen that when the RIS is closed to B, the secrecy rate of the legitimate system is reduce the most. 

\section{Conclusions}
This paper considers that an illegitimate receiver uses an RIS to help perform eavesdropping and/or jamming. We formulate and then solve the problems that minimize the secrecy rate of the legitimate system. Our simulations illustrate that eavesdropping performance can be enhanced by using the RIS. When RIS is further used to also perform jamming, the secrecy rate of the legitimate system is largely reduced. For eavesdropping and/or jamming, better performance of the illegitimate receiver can be expected by placing the RIS at an appropriate  position.

\ifCLASSOPTIONcaptionsoff
\newpage
\fi


%





\end{document}